\documentclass[11pt]{article} 

\usepackage{graphicx} 

\title{Measurements of Gravitational Attractions at small Accelerations}

\begin{document}

\author{%
W. Bartel\textsuperscript{1\dag}  C. W. Elvers\textsuperscript{1}  L. Jönsson\textsuperscript{2} G. Kempf\textsuperscript{1} 
H. Krause\textsuperscript{3}   \\                
 B.~Loehr\textsuperscript{1}  E. Lohrmann\textsuperscript{3},  H. Meyer\textsuperscript{4\dag}  
P. Steffen\textsuperscript{1$\star$} E. Wuensch\textsuperscript{1} \\
\\
\\
\textsuperscript{1}\normalsize Deutsches Elektronen-Synchrotron DESY, Hamburg, Germany \\
\textsuperscript{2}\normalsize Lund University, Sweden \\
\textsuperscript{3}\normalsize Universität Hamburg \\
\textsuperscript{4}\normalsize Bergische Universität, Wuppertal \\
\textsuperscript{\dag}\normalsize deceased \\
\textsuperscript{$\star$}\normalsize corresponding author: peter.steffen@desy.de \\   
}

\date{\  }

\maketitle

\vspace{4cm}
\begin{abstract}
Gravitational interactions were studied  by measuring  the influence of small external field masses
on a  microwave  resonator.
It consisted of two spherical  mirrors, which acted as independent pendulums
individually suspended by strings.  
Two identical field masses were
moved along the axis of the resonator symmetrically and periodically between
a near and a far position. 
Their gravitational interaction altered the distance between the mirrors,
 changing  the resonance frequency,
which was measured and found consistent with Newton's law of gravity.
The acceleration of a single mirror caused by the two field masses at the closest position  varied from 
 $\rm 5.4\cdot 10^{-12}\ m/ s^2$ to  $\rm259 \cdot 10^{-12}\ m/ s^2$.
\end{abstract}

 \newpage 
\section{Introduction}

In the 1930s, observations were made by several astronomers that the  galaxies contain more mass than expected from Newton's gravitational law, based on the visible mass.
The first publication of such an observation was made by K. Lundmark \cite{Lundmark}.
Later the term ``dark matter'' was coined. 
 The present predominant hypothesis is that dark matter is some kind of massive, stable, neutral, weakly interacting particle. 
 However, in spite of many researches to find or identify the  particles of dark matter, 
 nothing has been found so far \cite{gL_004b_L_Baudis_2018}, \cite{gL_034_DM_Overview_2023}.
An alternative to dark matter is the MOND-model that assumes  a  modified Newton’s
law of gravity \cite{gL_012_M.Milgrom_1983}.

A high statistics dataset has been evaluated by S. S. McGaugh et al.   \cite{McGaugh},  \cite{McGaugh2}.
In the acceleration range of about  $\rm (20  - 500)\ 10^{-12}\ m/s^2$  there is a clear deviation of  the measured  data from expectations 
obtained from  Newton's predictions based on the baryonic matter.
This experiment aims to test Newton’s gravitational law at accelerations below
  $\rm 10^{-10} \ m/s^2$ on Earth, relying solely on gravitational interactions.
  
 Small field  masses were used to accelerate the masses of two pendulums 
  perpendicular to the Earth's acceleration. The measured  movements of the pendulums 
  are proportional to the acceleration caused  by the field masses.

Results from measurements with field  masses ranging from 9 kg to  1 kg have been 
published previously \cite{gL_042_GRAVI_2011} using an earlier version of the experiment.
The results did not show any deviation from  Newton's law of gravitation. 
A similar result was obtained using an experimental  setup  of the Cavendish-type
\cite{Little}.

This paper presents  results from an improved experimental setup using field masses from 3 kg to 0.1 kg, 
which extend the range of accelerations to significantly smaller values,
measured with considerably higher precision.
The measurements were carried out in the years  2020 to 2022.

\section{Experimental Setup}

\begin{figure}[h]
\begin{center}
\vspace{-0.5cm}
\includegraphics[width=1.0\textwidth,height=0.35\textheight]
{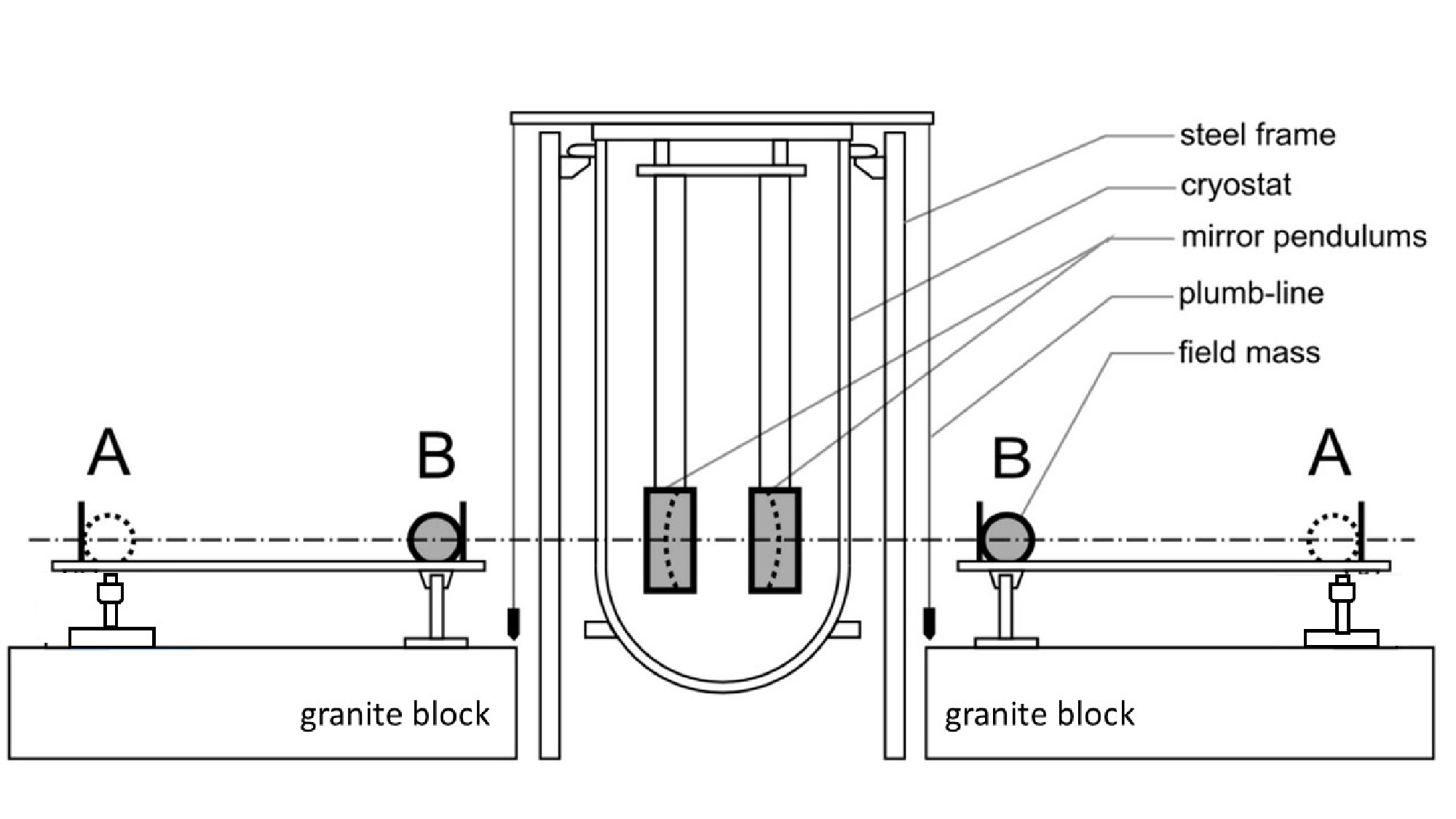}

\caption{Detector scheme: two mirror pendulums which are  pulled apart 
 by the gravitational forces of field masses on both sides. 
 The field masses change their positions between  A and B.
}
\label{fig:detector}
\end{center}
\end{figure}

Figure \ref{fig:detector} shows the main components of the experiment:
 a system of two pendulums that form a microwave resonator, similar to a Fabry-Perot interferometer.
The introduction of external field masses causes a tiny acceleration of the pendulums
in the direction of the resonator axis.		   
 The resulting change of the resonator length caused a  measurable  change
  in the resonance frequency.
 
 The resonator was located inside a vacuum vessel.
 A transport system moved the field masses between positions A and B.

An earlier version of the apparatus had been built and operated at Wuppertal University 
for a precision measurement of the gravitational constant: 
\cite{gL_013_Schurr_1992},
\cite{gL_014_Walesch_1995},
\cite{gL_015_Schuhmacher_1999},
\cite{gL_016_Kleinevoss_2001}, where details can be found.	
It was later transferred to DESY to an experimental hall of the PETRA accelerator.

	After an initial measurement period, which resulted in a publication 
	\cite{gL_042_GRAVI_2011},	
	it was relocated to an underground experimental hall of the HERA accelerator,  
	and the setup was  improved significantly.

\subsection{The Resonator}
\label{sec:reson}

\begin{figure}[h]
\begin{center}
\vspace{-0.5cm}
\includegraphics[width=1.0\textwidth,height=0.3\textheight]
{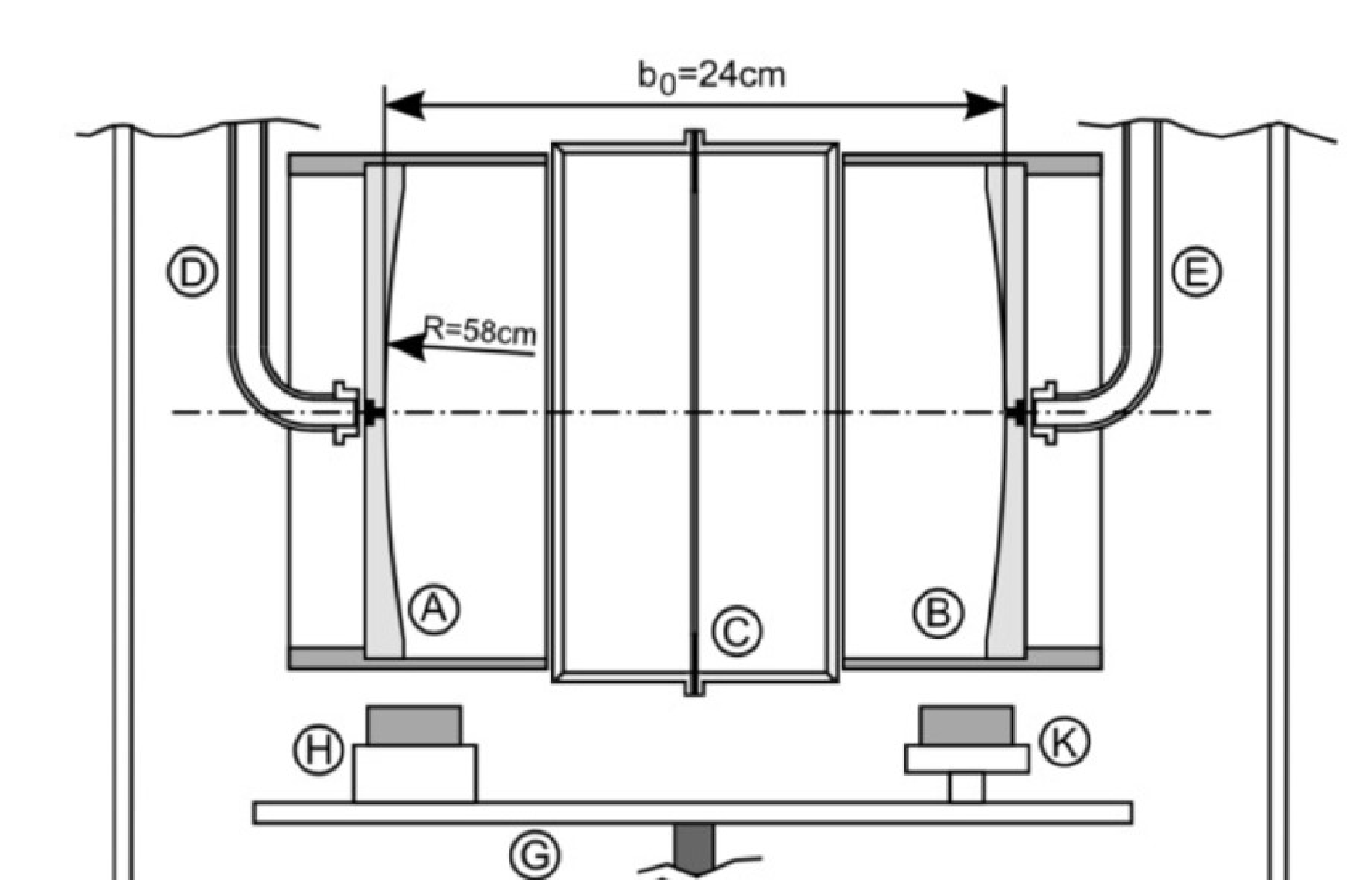}

\caption{Resonator scheme:  the two cylindrical mirror pendulums (A, B)  form a microwave resonator
together with a cylindrical mode filter (C) in between.
The mode filter suppresses higher mode resonances.
The radio frequency power is fed into the resonator  via the wave guide (D) .
The resonator  amplitude is measured at the opposite side (E).
Oscillations of the pendulums are damped by independently adjustable magnets (H, K)
on a movable table (G).
}
\label{fig:resonator}
\end{center}
\end{figure}

The central part of the experiment was a microwave resonator. 
Figure \ref{fig:resonator} shows a schematic view.
It consisted of two mirrors with spherical surfaces and 
a cylindrical center piece in between, which suppresses higher order resonance modes. 
The mirrors are made of copper with a gold plated surface on the concave side. They were enclosed by an aluminum tubus.
The mirrors and the center piece were suspended individually by thin strings
of about 2.7 m in length from the upper lid of the vacuum vessel.
The thin strings had a diameter of  0.2\  mm  and were made of a high density polyethylene (Dyneema). 
Each mirror acted as a mechanical pendulum with a period of 3.278\ s. 
The resonator was operated at a mean resonance frequency of 23.2699\  GHz. It was adjusted continuously to compensate slow drifts caused by changes of temperature and air pressure. 

Stochastic movements of the pendulums were damped by permanent magnets positioned below the mirrors
in order to optimize the performance of the resonator.
The  center of gravity of each mirror  coincided  with the vertex of the mirror.
The distance between the two vertices was determined to $\rm b_0 = 0.2400\ m$. 
Radio frequency power was fed into the resonator through 
a small hole in the center of one mirror, 
and  was similarly extracted on the opposite  side. 
It was rectified and read out.

\subsection{The Vacuum Vessel}
\label{sec:vacves}

In figure \ref{fig:detector}  a cryostat is shown. It was used as vacuum vessel  at room temperature.
It was a cylindrical stainless-steel tank with a removable lid on top. 
It was supported at the upper surface from a massive concrete structure.
Ground vibrations were suppressed by a mechanical damping system.

The vacuum vessel was thermally shielded with a layer of Styrofoam. 
A removable, thermally insulating hood covered the top of the concrete structure,
maintaining the lid's temperature variation at about 0.1 degrees Celsius within a  12-hour period. 
Over longer time periods the temperature changes were about a factor of 10 higher, 
especially between summer and winter. 
 At the height of the resonator the vacuum vessel  was shielded against magnetic fields by a 
$\rm\mu$-metal cylinder. 
The lid of the vessel was leveled along and transverse to the resonator axis 
with a precision of a few micro rad.
Variations of the lid’s inclination occurred up to $\rm 10\ \mu rad$. 

Environmental  changes of air pressure and temperature led to a drift of the resonance frequency of up to 10 kHz within a  week.

\subsection{The Transport System of the Field Masses}

Two granite optical benches were located on  opposite sides of the vacuum vessel along the centerline of the resonator. 
Both benches supported  a transport system on which  field masses were moved 
 between a far and a near position (see figure \ref{fig:detector}).
The distance between the two positions was always 2.683 m on either side.

The construction of the transport system allowed for  a change in  the near position on both benches.
This flexibility enabled  different gravitational accelerations on the mirrors with given field masses 
(for details, see section \ref{sec:FM}). 
Several spheres of different materials but with the same diameter of 0.127 m were used as field masses.

\subsection{Improvements}
\label{sec:improve}

After the  relocation   to an underground experimental hall of the HERA accelerator 
the experimental setup was significantly improved:
\begin{itemize}
\item[] The lid inclination was stabilizes by using a damped suspension of the vacuum vessel.
\item[]  Temperature stability of the lid was maintained down to 0.1 degrees Celsius over several hours.
\item[]  Stable vacuum conditions were maintained  in the vessel at about  $\rm 2 \cdot 10^{-5}\  mbar$.
\item[]  The method  for determining  the near positions of the field masses was improved,
\item[]  The data acquisition was significantly improved by the installation of a recursive algorithm for the continuous readjustment to a drifting resonance frequency.
This resulted in a  more stable data collection process independent of external large distortions, like earthquakes.
Further on, it allowed to choose a resonance frequency with a narrower width resulting in an improved frequency resolution.
\end{itemize}

\section{Predictions from Newton’s Law of Gravitation}
\label{sec:pred}

The size of the deflection of a pendulum is determined by the equilibrium between
the  gravitational attraction by the field masses  and the restoring force  of  the Earth's gravitation.

For point masses, the gravitational force F between a field mass m and a pendulum with mass M 
at a distance of r  is given by:
\begin{equation}
\rm F = G \cdot \frac{m}{r^2}\cdot M
\end{equation}
The acceleration of the pendulum by the field mass  is 
\begin{equation}
\rm a(m,r) =  G \cdot \frac{m}{r^2}
\end{equation}

Two equal field masses 
were positioned symmetrically w.r.t. the resonator.
They were  alternated periodically between a near ($\rm r_n$) and a far  ($\rm r_f$) position.
Assuming point masses for field masses and mirrors,
the change of the resonator length between these two positions 
is described by the formula

\begin{equation}
\rm db_{point} = 2\ \frac{G}{\omega_0^2}\cdot m\cdot
 \bigg\lgroup\big\lgroup \frac{1}{r_{n}^{\,2}}  -\frac{1}{({r_n +b)^2}}\big\rgroup  
-  \big\lgroup\frac{1}{r_{f}^{\,2}}- \frac{1}{({r_f+b)^2}} \big\rgroup\bigg\rgroup
\end{equation}

where $\rm \omega_0^2 = g/L$ is the frequency of a pendulum with
the length $\rm L$ and $\rm g$ is the local gravitational field of Earth.
b is the distance of the vertices as well as the centers of gravity of the two mirrors.

In terms of acceleration one has 

\begin{equation}
\label{pointacc}
\rm db_{point} = \frac{2}{\omega_0^2}\bigg\lgroup\big\lgroup a(m,r_n) - a(m,r_n+b)\big\rgroup 
- \big\lgroup  a(m,r_f) - a(m,r_f+b)\big\rgroup\bigg\rgroup \label{eq:dbpoint}
\end{equation}

The field masses were homogeneous spheres, treated as point masses.
However, the mirrors have a complicated geometric structure composed of different materials.
The acceleration of the extended mirrors were calculated by numerical  integration over the shapes of different densities:
\begin{equation}
\rm a(m,r)_{ext} = G\cdot  \frac{1}{M}\cdot \int_V \frac{\rho(V)\cdot m}{r(V)^2} dV ,
\end{equation}
where $\rm \rho(V) $ is the density of the volume elements, and 
 r(V) is the distance of field mass and volume element projected onto the resonator axis.
 
Using $\rm a(m,r)_{ext}$ in formula \ref{eq:dbpoint} gives the predicted result:

\begin{eqnarray}
\rm db_{pred} & = & \frac{2}{\omega_0^2}\bigg\lgroup\big\lgroup a(m,r_n)_{ext} - a(m,r_n+b)_{ext}\big\rgroup  \nonumber \\
  & - &\ \ \ \ \   \big\lgroup  a(m,r_f)_{ext} - a(m,r_f+b)_{ext}\big\rgroup\bigg\rgroup  \label{eq:dbpred}
\end{eqnarray}

The ratio of $\rm db_{pred}$  for extended and point masses  ranges from 0.96 to 0.99.
The contribution of the far position to $\rm db_{pred}$ is about 1\%.
It was neglected.
Therefore  the relevant acceleration on a single mirror by the two field masses in the near position
is described by 
\begin{eqnarray}
\rm a_{pred} = \frac{g}{L} \cdot \frac{db_{pred}}{2} \label{eq:apred}  \mbox{\ \ \ \  and  \ \  \ \ }
\rm a_{meas} = \frac{g}{L} \cdot \frac{db_{meas}}{2} \label{eq:ameas} , \mbox{\ \ \ \  resp. }  \label{eq:acc}
\end{eqnarray}

For the actual calculations the following values were used:
\begin{eqnarray*}
 \rm  G & = & \rm 66.7428 \cdot 10^{-12}\ \frac{ m^3}{  kg \  s^2} \ \ \ \ \ \     \cite{gL_031_G_Tiesinger_2021}  \\		
 \rm  g & = & \rm 9.8138\ m / s^2	\ \ \ \ \ \ \ \ \ \ \ \ \ \ \ \ \ \ \cite{gL_044_GRAVI_2016}	
\end{eqnarray*}

\section{Measuring Procedure}

The aim of the experiment is to measure the gravitational effect of the two field masses
on the mirrors of the resonator.
This is determined from the resonance frequencies of near and far position data.

\subsection{Frequencies of the Resonator}
\label{sec:freq}

According to \cite{gL_12_Kwai-Man_1985},  
the resonance frequencies of a cylindrical resonator with spherical mirrors are given by:
 
\begin{equation}
 \rm f = \frac{c}{2b}\cdot \Bigg\{ q + \frac{n}{\pi}arcos\bigg(1 - \frac{b}{R}\bigg) + \frac{N}{8\pi ^2 R k}
  + \mathcal{O}\bigg(10^{-4}\bigg)   \Bigg\}    \label{equ:fres}
\end{equation}
where:
\begin{itemize}
\item[]  q = axial mode number.
\item[]  p = radial mode number.
\item[]  m = azimuthal mode number.
\item[]  $\rm N = 2p^2 + 2pm - m^2 + 2p - 2 + m \pm 4m$.
\item[]  $\rm n = 2p + m +1$.
\item[]  R = 0.56 m, the  curvature radius of spherical mirrors.
\item[]  c the speed of light.  
\item[]  k  = f / c the wave number.
\end{itemize}
Our experiment was performed at a resonance frequency with p = m = 0. 
The parameters q, p, m have been determined by a simultaneous fit of   formula \ref{equ:fres} 
to a series of resonances including the chosen one,
where the  axial mode number is 37.

The relation between db and its corresponding change in resonance frequency 
df is given by:
\begin{equation} 
\rm  db = \beta\cdot df\   \mbox{\ \ \  with}    \label{eq:db}
\end{equation}
\begin{equation}
 \rm \beta = -\frac{b}{f} \Bigg( 1 - \frac{n\cdot c}{2\pi\cdot f}\sqrt{\frac{1}{2 R b - b^2}} 
+  \mathcal{O}\bigg(10^{-4}\bigg)\Bigg). \label{equ:beta}
\end{equation}

For the used resonance frequency of \mbox{ 23.2699\ GHz} 
it results in $\rm \beta = 10.371$   $10^{-12}\  m/Hz$.
$ \rm \beta$ is important for the conversion of the theoretically expected  db  to the  expected df, 
the frequency difference expected for data with near and far field mass.
It allows a direct comparison of predicted and measured values of df.
This comparison  requires an evaluation  of the uncertainties of the used parameters.
$ \rm \beta$   depends essentially on the parameters   f and b.
The relative uncertainty of f, $ \rm \delta f / f$, was determined from the slow drift of the resonance frequencies 
around the  average resonance frequency of the  whole 2 1/2 year measuring period.
It amounts to $ \rm \delta f / f < 0.2 \  10^{-6}$.
According to equation  \ref{equ:fres}, the relative uncertainty of b is  $ \rm \delta b/b = -\delta f/f$ .
Therefore the  relative uncertainty of $ \rm \beta$  is $\rm  \delta \beta /  \beta  \approx 2\cdot \delta f  / f  <   0.4 \  10^{-6}$.

These uncertainties are very small as compared to the results of measurements and theoretical predictions (see table \ref{tab:DFpred}).
They were neglected in the numerical calculations of the theoretical predictions,
where $\rm b = b_0$  and f = 23.2699\ GHz were assumed.

\subsection{Determination of the Resonance Frequency}
\label{sec:resfrq}

The resonance frequency, $\rm f_r$,  was  determined from measurements of the resonator amplitudes
at  five   equidistant frequencies (10 kHz difference). These were adjusted continuously to center approximately around the slowly drifting resonance. 
These frequencies cover 30\% of the resonance width at half maximum of the resonance shape (about 130 kHz).

A Lorentz curve is expected to describe the resonance shape:
\begin{equation}
\rm   U(f_i)  = U_{max} \cdot \frac{1}{1 + 4((f_i - f_r)/f_w)^2} \ \ \ \mbox{for i = 1 - 5}  
\label{equ:Lorentz}
\end{equation}
The maximum amplitude  $\rm U_{max}$, the width of the Lorentz curve $\rm f_w$ , and
the resonance frequency  $\rm f_r$ could be  obtained from a fit of equation \ref{equ:Lorentz}
to the 5 amplitude measurements $\rm  U(f_i) $.

However,  the inverted Lorentz curve is a parabola that was used for a  much faster fit:
\begin{eqnarray}
\rm   U(f_i)^{-1} &=&  a + b\cdot f_i + c\cdot f_i^2 \ \ \ \mbox{for i = 1 - 5 ,   \ \ \  with} \\ 
\rm f_r &=&  - \frac{b}{2c}
\end{eqnarray}
The parabola fit  was sufficiently  fast for an online determination, and
allowed an immediate adjustments of the measurement range to a drifting resonance frequency.

The resolution $\rm \sigma(U(f_i))$ was about $\rm 0.05\ mV$.
The resolution of the resonance frequency  benefited  from the improved data acquisition system 
(see section \ref{sec:improve}).

\subsection{ Field Masses}
\label{sec:FM}

The used field masses were spheres of different materials and mass values, each 
with a diameter of 0.127 m, as specified in the table below.

\begin{center}
\begin{tabular}{|r|r|r|r|r|}
\hline
   material  &    marble      &        plastic       &      plastic       &         plastic  \\
\hline
   mass   &      2.92 kg    &          1.002 kg   &         0.294 kg &          0.116 kg \\
\hline
   accuracy &      $\rm\pm\  0.1\ g $    &      $\rm\pm\  0.1\ g $    &      $\rm\pm\  0.1\ g $    &      $\rm\pm\  0.1\ g $    \\
\hline
  structure  &     solid  &                solid   &           hollow     &          hollow \\
\hline
 \end{tabular}
\end{center}    

Measurements were performed with different configurations of  field masses and
distances between the gravity centers of the  field mass and the closer mirror
as specified in the table below:
 \begin{center}
\begin{tabular}{|r|r|r|r|r|r|r|r|}
\hline
 mass   [kg]   & 0.116   &0.294   &0.116   &0.294   &0.294   &1.002   &2.924    \\
 \hline
  $\rm r_n$  [m] & 0.7537  &1.0127   &0.5932   &0.7537   &0.5932   &0.5932  &0.5932  \\
\hline
\end{tabular}
\end{center}  
The positions of the two field masses were changed every   half an hour 
from their far positions   to their near positions and vice versa.  
About 5000 measurements of the resonance frequency were registered per half hour period.
The  duration of data collection for each configuration ranged from 500 to  1000 hours.

\section{Systematic Effects}
\label{sec:syst}

Two categories of systematic uncertainties affect the comparison of the results 
with the expectations from Newton’s Law of gravitation. 

The first category consists of effects that might or do influence the determination of the resonance frequency:
\begin{itemize}
\item[]  The frequency  stability of the radio frequency  generator output was compared to a Rubidium Standard frequency. 
No difference was found that would influence the resonator frequency determinations. 
\item[]  The removable cover of the vacuum vessel behaved like  a membrane. 
External influences, such as  temperature and pressure, led to changes of the resonance frequency. 
In section \ref{sec:datana} it is described how this was treated in the analyses of the data.
\item[]  Measurements  with no movement of the field masses were performed to test whether there is a bias in the measurement procedure and the analysis methods.  
The result  $\rm df = -0.058 \pm 0.063 \ Hz$ is compatible with zero. 
This result was obtained following the analysis procedures  described in section \ref{sec:analyse}.
\end{itemize}

The second category of uncertainties  is associated with the calculation of expectations 	from Newton’s law of gravity.
\begin{itemize}
\item[]
 A possible  asymmetry of the resonator position in the experimental setup
 was determined using two methods:
 \begin{itemize}
\item[]
Measurements  starting with one field masses in the near position 
 and the other in the far position.
 These positions were exchanged  every 0.5 h.
 \item[]
Measurements with only one single field mass on one side,
 alternating between the far and near position.
 A second set of measurements were performed with the same field mass 
 on the opposite side.
\end{itemize}

 Both methods resulted  in an asymmetric position of the resonator of about 
 $\rm 5\ mm$. 
The asymmetry effect almost completely cancels  when two  field masses are used.
 The remaining deviation amounts to a negligible  0.03\ \%  change of df.

\item[] The distances between the centers of gravity of the field masses to the centers of gravity of the
mirrors  play a critical role  in the calculation of the expectation from Newton’s law. 
The distances in the near position could not be measured directly because the resonator was located in the vacuum vessel between the field-masses.
Instead, the distance, D, between the centers of the two field masses in the near position was measured to be 
$\rm D = 1.4270 \pm  0.0005\ m$, resulting in
 \begin{equation}
 \rm r_n = \frac{1}{2} (D - b_0) =  0.5935\  m\ .
 \end{equation}
  
\item[] The difference between the near and far position of the field masses was always 
 \mbox{$\rm  2.683 \pm 0.003 \ m$. }
The contribution of the masses in the far position was about 1\ \% of the total df.
Therefore the uncertainty of the far position is negligible.

\item[] The field masses were measured with an electronic scale with an uncertainty of 0.1 g. 
Small differences between the  field mass pairs cancel  completely in the calculation of the prediction of df
if the average of the two field masses are used.

\item[] The horizontal and vertical alignments of the field masses transport systems at the near positions deviate by at most 2 mm from the axis of the resonator. This has a negligible effect on the determination of df.

\item[] The accuracy of  the predicted $\rm db_{pred}$ depends, according to equation \ref{eq:dbpred}, on $\rm 1/\omega_0^2 = L/g$, 
which depends on L,, the pendulum length.
 This length has been measured to be $\rm 2.671  \pm 0.001\ m$.
 The resulting relative uncertainty in the calculation of $\rm db_{pred}$ can be neglected.

\end{itemize}

\section{Raw Data }
\label{sc:raw}

\begin{figure}[h]
\begin{center}
\vspace{-1.5cm}
\includegraphics[width=1.0\textwidth,height=0.30\textheight]
	{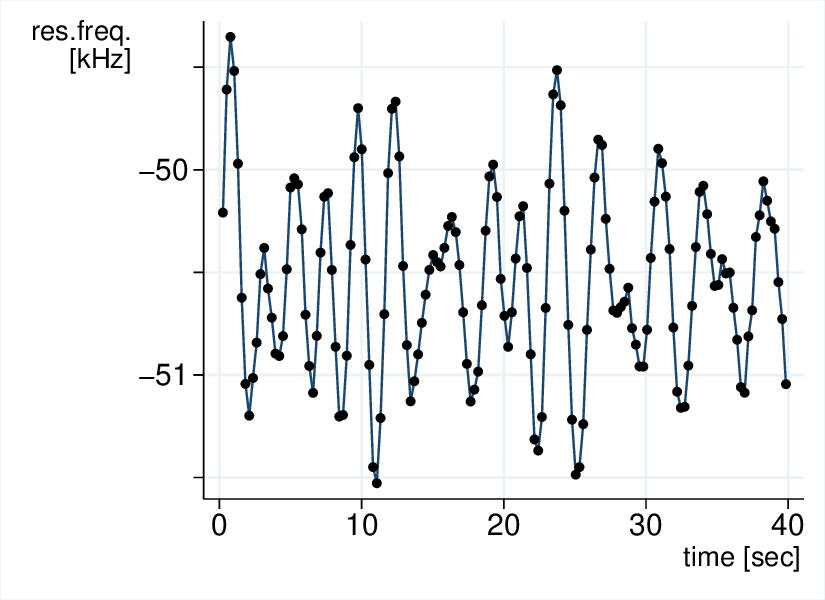}

\caption{The points show  the  resonance frequencies for a time period of about 40 sec,
after a fixed offset was subtracted.
They were connected by straight lines 
to demonstrate clearly the effect of pendulum oscillations.
 }

\label{fig:rawfreq2}
\end{center}
\end{figure}

The mirrors  are permanently excited by vibrations of the ground floor resulting in oscillations 
of the resonance frequency. 
These vibrations  added to  the gravitational  effects of the field masses in near and far positions.
Figure \ref{fig:rawfreq2}  shows the originally determined resonance frequencies over 
a time interval of 40 s. 
The frequencies given on the y-axis are obtained after subtraction of a constant offset.
Single frequencies  are shown as dots.
	
The bandwidth of the resonance frequencies is primarily dominated
by pendulum oscillations, resulting from relative distance changes between the two mirrors.
The pendulum period of approximately 3\ s is  clearly visible, 
as well as the data taking period of about  0.3 s.
The resonance frequencies were filtered by averaging them over 30 to 60 second intervals 
(10 - 20 pendulum periods).	
During these intervals, the pendulum effect averages  out.

\begin{figure}[h]
\begin{center}
\vspace{-0.5cm}
\includegraphics[width=1.0\textwidth,height=0.30\textheight]
	{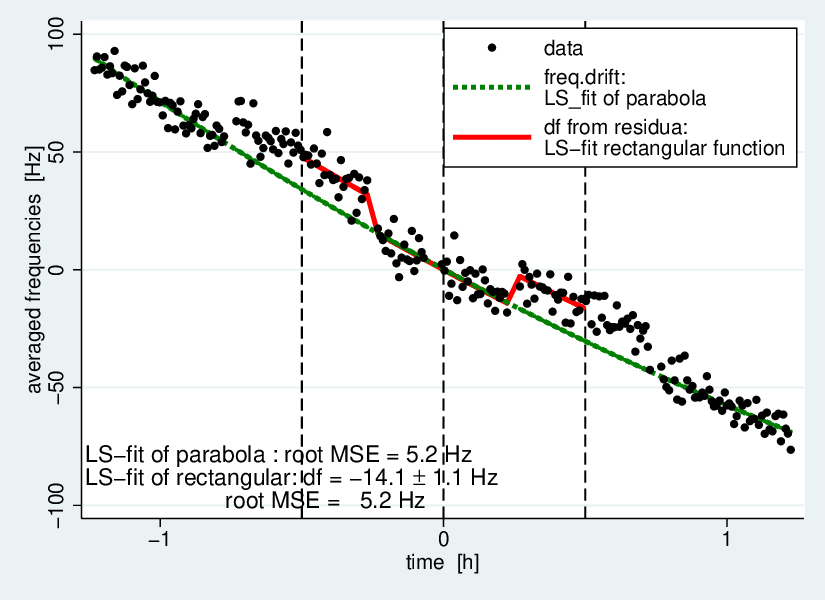}

\caption{The points show  the means of the resonance frequencies, averaged over 30 sec,
for a time interval  of 2.5 h.
A fixed offset was subtracted from the frequencies as well as for the horizontal time axis.
The data were derived from measurements with 2.924 kg field masses. 
A slow drift of the  frequency of about 150 Hz is observed within the 2.5 h time interval, as described by the green line.
Every half hour, a frequency step is observed originating from the position change of the field masses, described by the red line.
 }

\label{fig:df_det}
\end{center}
\end{figure}

The left-over averaged resonance frequencies are shown as black dots in figure  \ref{fig:df_det} over a 2.5 hour span.
They show a slow  drift of the resonance frequency (see section \ref{sec:vacves}).
A frequency step occurs between half-hour intervals  when the field masses alternate between near and far positions.
The frequency fluctuations form a band with a standard deviation of about 5 Hz around the drifting resonance frequency.

Only the averaged and offset subtracted resonance frequencies were used in the further analysis of the data.

\section{Data Analyses}
\label{sec:analyse}

Four data analysis methods were developed separately.
In general the  least square method  was used for all fits,  unless otherwise stated.

\subsection{Different Analysis Methods}
\label{sec:datana}

Independent methods (m = a,b,c,d) were developed for the analyses of the data.
All analyses begin with the data selection: 
\newline Strong temporary distortions like earthquakes etc.  were eliminated by 
visual inspection \mbox{(a, b)} or by a programmed procedure (c, d).

The  methods used time intervals of 1 to 10 hours  
for the determination of  initial results of df  and their statistical uncertainties,
taking into account the slow frequency drift. 

\begin{figure}[h]
\begin{center}
\vspace{-0.5cm}
\includegraphics[width=1.0\textwidth,height=0.30\textheight]
	{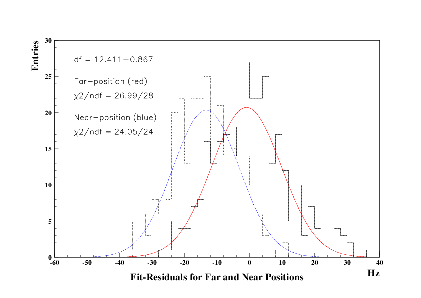}

\caption{Shown are two histograms of the residuals of  a 5th order polynomial fit to a 
selected 10-hour data-taking period, one for the field masses in the near position(dashed histogram),
one for the far position(solid histogram).
Gaussian curves were fitted to the histograms in blue and red. 
The resulting values of  $\rm \chi^2 / ndf$  are close to one,
demonstrating the goodness of the Gaussian fit.
These also demonstrate  the accurate description of the slow frequency drift by the 5th order polynomial fit.
 }

\label{fig:dfresBL}
\end{center}
\end{figure}

The  different methods are:
\begin{itemize}
\item[] Method a:
Time intervals of 10 hours were used.
The slow frequency drift is described by a 5th order polynomial, 
that was fit to the frequencies of the periods with field masses in the far position.
 The residuals  of the fit for near and far periods 
 are  two nearly Gaussian distributions:
 one for the near position data and one for the far position data.
  df is the difference of the two means. 
  The uncertainty of df follows from the statistical uncertainty of the two means.
  
  As an example,  figure \ref{fig:dfresBL} shows distributions of the residuals of the 5th order polynomial fit 
  to a selected 10-hour data-taking period, 
  one for the field masses in the near position(dotted histogram) and one for the far position (solid histogram).
The Gaussian fits of the histograms resulted in  $\rm \chi^2 / ndf$   close to one.
These demonstrate also the accurate description of the slow drift by the 5th order polynomial fit.
  
\item[] Method b: 
A parabola  plus a rectangular function was fit to the resonance frequencies
for  3 consecutive periods.
The fit step-size  of the rectangular function df and  its statistical uncertainty were obtained from the fit.
\item[] Method c:
As shown in figure  \ref{fig:df_det}, a parabola  was fit to the 
frequencies of 3 consecutive  periods with the same position of the field masses.
The green line of the parabola describes the slow frequency drift with sufficient accuracy.
The values of the parabola have been subtracted from the data.
A  rectangular function were fit to the resulting residuals  of 1 hours data giving the step-size df and its statistical uncertainty.
The fit result is shown as red line on top of the parabola in the figure.
\item[] Method d: 
A 5th order polynomial plus a  rectangular wave were fit to all frequencies
of a time intervals of 10 hours.
The parameters of the polynomial describe the slow frequency drift.
The rectangular wave  changes its sign at every alternation between near and far position.
The average step-size of the rectangular wave df 
and its statistical uncertainty was obtained from the fit.
\end{itemize}
For each method the initial results were combined  to averages  $\rm df_m \pm \sigma df_m$
for each  configuration of field masses and near distance.
Table  \ref{tab:DFcomb} shows the results of the different methods in columns 2 to 5
together with the predicted acceleration in column 1.

\subsection{Combination of the four Methods}

The  mean value $\rm \overline{df}$ of the methods as well as its statistical uncertainty  $\rm  \sigma_{stat}$
were computed as weighted averages of the $\rm  df_m$  and the  $\rm  \sigma df_m$:  
\begin{equation}
  \rm \overline{df} =\bigg( \sum_m df_m \cdot (\sigma df_m)^{-2} \bigg) \bigg/ \bigg(\sum_m ( \sigma df_m)^{-2} \bigg)
\end{equation}
\begin{equation}
 \rm \sigma_{stat} =\bigg(\sum_m \sigma df_m \cdot (\sigma df_m)^{-2}  \bigg) \bigg/  \bigg(\sum_m( \sigma df_m)^{-2} \bigg)
\end{equation}

\begin{figure}[h]
\begin{center}
\vspace{-0.5cm}
\includegraphics[width=1.0\textwidth,height=0.38\textheight]
{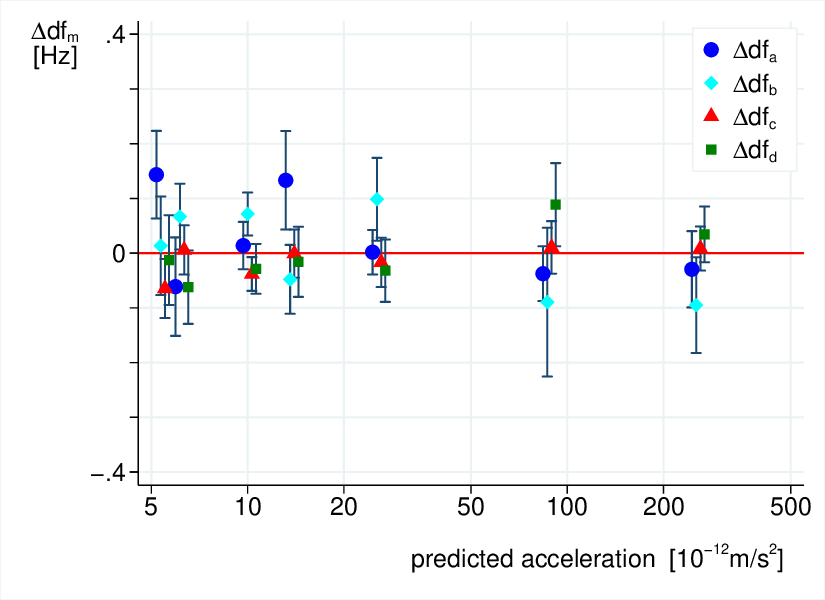}

\caption{The differences 
 \mbox{$\rm\Delta df_m = (df_m \pm \delta df_m)  - \overline{df}$}
of the different  methods are shown for the predicted accelerations. 
They are presented  slightly displaced for better visibility.
}
\label{fig:Ddfm}
\end{center}
\end{figure}

Figure \ref{fig:Ddfm} shows the differences  $\rm (df_m \pm \sigma df_m) - \rm \overline{df}$
for the different accelerations,
demonstrating  good agreement of the results from different analysis methods.

The standard deviation of $\rm \overline{df}$ describes 
the spread of the method results $\rm d f_m$ around the mean  $\rm \overline{df}$.
It  is used  as additional systematic uncertainty, $\rm\sigma_{sys}$,
for the  difference between the four  methods.
Table \ref{tab:DFcomb} presents  $\rm \overline{df}$, 
as well as $\rm\sigma_{stat}$ and $\rm\sigma_{sys}$  in columns 6.

The combined uncertainty is 
\begin{equation}
\rm \delta\overline{df}  = \sqrt{\sigma_{stat}^2 + \sigma_{sys}^2 }.
\end{equation}

\section{Results}

Table \ref{tab:DFpred} gives  $\rm\overline{df} \pm \delta\overline{df}$ 
 in column 1 for the different accelerations.
 The results for $\rm \overline{df}$  have uncertainties that are
 nearly the same  for the different accelerations:
the mean uncertainty is $\rm \overline{\delta \overline{df}} = 0.074\ Hz$.
  At the lowest acceleration
 $\rm \overline{df}$ could be determined with a significance of  
 $\rm 3.5 \cdot \delta\overline{df}$.
 Therefore measurements at lower acceleration would have given only insignificant results.

The predictions of Newton's law, $\rm df_N \pm  \delta df_N$, are displayed in table  \ref{tab:DFpred}  column  3.
The uncertainties  $\rm\delta df_N$ were determined from 
the uncertainty of $\rm r_n$.
 
\begin{figure}[h]
\begin{center}
\vspace{-0.5cm}
\includegraphics[width=1.0\textwidth,height=0.38\textheight]
	{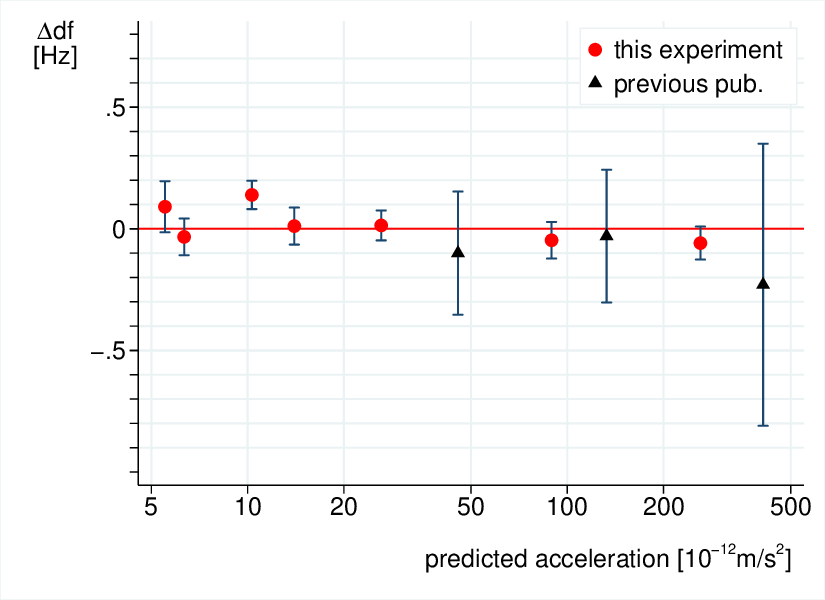}

\caption{Differences, $\rm \Delta df = \overline{df} - df_N$,  of average results  and predictions
are shown  for the predicted accelerations.
The vertical error bars include the uncertainties of measurement and prediction.
The red symbols show  the results of this experiment.
The black symbols show the results derived from  the previous publication.}
\label{fig:DDF}
\end{center}
\end{figure}

The differences of measurement results and predictions, $\rm \Delta df \pm \delta \Delta df$, 
are given in table \ref{tab:DFpred}  columns 5 for the different accelerations.
Measurements and predictions agree  well within the uncertainties.
This is shown in in figure \ref{fig:DDF}  by the red symbols.

Also shown, as black symbols,  are the results
  of an earlier publication  \cite{gL_042_GRAVI_2011}.
These were determined from the information 
taken from \cite{gL_011a_S_Schubert_2011} on which the publication was based.
It demonstrates the large improvement in accuracy and the extended acceleration range that has been obtained 
since the previous publication.

\begin{figure}[h]
\begin{center}
\vspace{-0.5cm}
\includegraphics[width=1.0\textwidth,height=0.60\textheight]
	{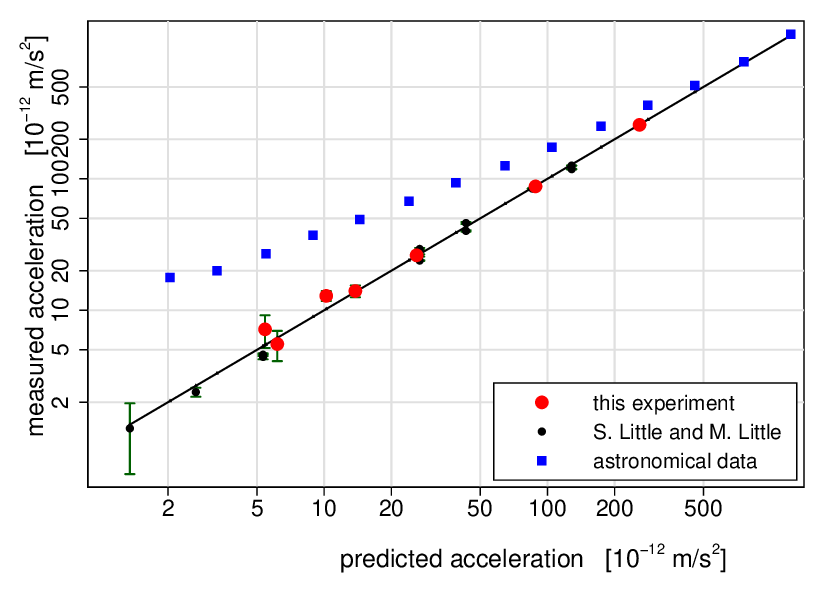}

\caption{Comparison of measured and predicted acceleration
of this experiment (red dots).
The relative uncertainty of the predicted acceleration is about 0.14\%.
The black line indicates the equality of the two accelerations.
The black symbols show the results from  a previous publication \cite{Little}.
The  results of this experiment are compared to astronomical measurements
\cite{McGaugh2} (blue squares);
vertical axis: radial accelerations by galaxies measured at the boundaries;
horizontal axis: predicted acceleration by the baryonic  mass according to Newton's law.
The uncertainties of the astronomical data are within the size of the symbols.}
\label{fig:accMcG}
\end{center}
\end{figure}

The accelerations on a single mirror were calculated 
using equations \ref{eq:acc} and  \ref{eq:db} for  the measured accelerations and the Newtonian predictions and  their corresponding uncertainties.
The  results are  shown in table  \ref{tab:DFpred} in columns 2 and 4.
In figure  \ref{fig:accMcG}  the measured accelerations are displayed versus the predictions
from Newton’s Law. The data are in good agreement with the predictions.
The latter are indicated by the solid line.
 Also shown are results from reference [7]. They
agree well with the data of this experiment. 
The numerical values plotted here are not given in [7] but were provided by the authors as private communication. 
For comparison figure \ref{fig:accMcG} contains in addition the astronomical
data provided by S. S. McGaugh  \cite{McGaugh},  \cite{McGaugh2}. For these data the predictions for the
expected accelerations are based on the known baryonic matter.

\section{Discussion}

The results of this experiment agree well with Newton’s Law for accelerations down to $\rm 10^{-12}\ m/s^2$.
This is not the case for astronomical data where a significant discrepancy is observed for accelerations  $\rm < 10^{-10}\ m/s^2$ 
as shown in \mbox{figure  \ref{fig:accMcG}.} 
This does not  indicate that measurements on Earth contradict models and theories 
which explain the astronomical data.
The discrepancy between measured and expected accelerations demonstrates that the baryonic matter alone is not sufficient
to describe the measured results.

Dark matter models explain data from regions in universe which are not dominated by
gravitational effects of baryonic matter. This is not the case on Earth where gravitational effects are
dominated by baryonic matter.
In  figure  \ref{fig:accMcG} the predicted accelerations for astronomical data are calculated for
baryonic matter only. Dark matter causes additional gravitational acceleration. In
the region of small baryonic acceleration, the presence of dark matter leads to
higher predictions for the accelerations and may lead to an agreement with
Newton’s Law.

The MOND model has been formulated for regions where the gravitational
potential is very small which practically eliminates external forces. This
condition is not fulfilled on Earth. Therefore, the results of this experiments can
neither prove nor disprove models and theories which are based on MOND.

\section {Acknowledgment}

We  thank the DESY Directorate and the IT-division for their constant support.
We are grateful for the assistance of the technical groups of DESY for their help and advice
 on many technical questions.
Additionally we acknowledge the contributions of  A. Brüdgam, S. Fleig, Y. Holler, T. Külper,  
S. Karstensen, and U. Packeiser in setting up the experiment.
We express our gratitude for the fruitful discussions with the late N. Klein, W. Buchmüller,
S. Glazov, C. Niebuhr,  M. Takahashi, K. Schmidt-Hoberg, and  A. Ringwald.
We thank  S. S. McGaugh for providing the astronomical data and 
S. Little and M. Little for providing the numerical values of their results.

\newpage
\begin{table}[h]
\begin{center}
\begin{tabular}{|r|r|r|r|r|l|}
\hline
    &          &          &          &          &  \\
$\rm a_{pred}$  & $\rm df_a\   [Hz] $  & $\rm df_b\   [Hz] $  & $\rm df_c\   [Hz] $  & $\rm df_d\   [Hz] $  & 
$\rm\overline{ df}\   [Hz] $ \ \  \\
 \  $\rm [10^{-12}\ m/s^2]$    &   $\rm\sigma df_a $\ \ \   &   $\rm\sigma df_b $\ \ \   &   $\rm\sigma df_c $\ \ \   &   $\rm\sigma df_d $\ \ \   & \   $\rm\sigma_{stat} \ \ \ \ \sigma_{syst} $\ \ \    \\
\hline
 5.4 & 0.520 & 0.390 & 0.312 & 0.364 &\   0.377   \\
     & 0.080 & 0.090 & 0.054 & 0.082 &\  0.070  \ \   0.078 \\
\hline
 6.2 & 0.230 & 0.358 & 0.297 & 0.229 &\  0.291   \\
    & 0.090 & 0.060 & 0.045 & 0.067 &\  0.058 \ \   0.048  \\
\hline
 10.2 & 0.691 & 0.749 & 0.639 & 0.648 &\  0.677   \\
    & 0.043 & 0.039 & 0.031 & 0.046 & \ 0.038   \ \ 0.044  \\
\hline
 13.8 & 0.870 & 0.689 & 0.736 & 0.721 &\  0.737   \\ 
    & 0.090 & 0.063 & 0.044 & 0.064 &\  0.058 \ \   0.050  \\
\hline
 25.9  & 1.380 & 1.477 & 1.362 & 1.347 & \ 1.379   \\
      & 0.041 & 0.076 & 0.045 & 0.057 &\  0.049  \ \   0.037 \\
\hline
 88.5 & 4.570 & 4.513 & 4.618 & 4.696 &\  4.608   \\
    & 0.050 & 0.136 & 0.048 & 0.076 &\  0.058    \ \ 0.048 \\
\hline
 258.6 & 13.512 & 13.447 & 13.550 & 13.576 & 13.542   \\ 
    & 0.070 & 0.088 & 0.040 & 0.051 & \ \ 0.052    \ \ 0.037  \\
\hline

\end{tabular}
\caption{Results and uncertainties of the 4 methods 
are listed in columns 2 to 5 for the different $\rm a_{pred}$.
The combined results with statistical and systematic uncertainties are listed in column 6.
}
\label{tab:DFcomb}
\end{center}

\end{table}

\vspace{-2.0cm}

\begin{table}[h]
\begin{center}
\vspace{-2.0cm}
\begin{tabular}{|r|r|r|r|r|}
\hline
           &          &        &    &      \\
  $\rm \overline{df} \ \  [Hz]$ & $\rm a_{meas}$ \ \ & $\rm df_N \ \  [Hz]$ & $\rm a_{pred}$ \ \ & $\rm\Delta df$ \ \ [Hz]  \\
   $\rm\delta\overline{df} $\ \ \   &  $\rm\delta a_{meas} $   &   $\rm\delta df_N $\ \ \ &  $\rm\delta a_{pred} $  &   $\rm\delta \Delta df $\ \ \      \\
 \hline
 0.377 & 7.2 & 0.286 & 5.4 & 0.091  \\
       0.105 &    2.0& 0.0004 & 0.0 & 0.105 \\  
  \hline 
 0.291 & 5.5 & 0.324 & 6.2 & -0.033  \\
   0.075 &   1.4 & 0.0004 & 0.0 & 0.075  \\
 \hline 
 0.677 & 12.9 & 0.538 & 10.2 & 0.139  \\
    0.058 &    1.1 & 0.0009 & 0.0 & 0.058  \\
 \hline 
 0.737 & 14.0 & 0.726 & 13.8 & 0.011 \\
      0.076 &   1.4 & 0.0010 & 0.0 & 0.076  \\
    \hline 
 1.379 & 26.2& 1.365 &25.9 & 0.014  \\ 
    0.061 &  1.2 & 0.0022 & 0.0 & 0.061  \\ 
\hline 
 4.608 & 87.6 & 4.655 & 88.5& -0.047  \\ 
     0.075  &   1.4 & 0.0076 & 0.0 & 0.075  \\ 
 \hline 
  13.542 & 257.5 & 13.600 & 258.6 & -0.059  \\ 
     0.064 &   1.2 & 0.0211 & 0.0 & 0.068  \\
 \hline 

\end{tabular}
\caption{The combined results  for $\rm \overline{df}$ and the corresponding accelerations, $\rm a_{meas}$, 
the prediction, $\rm df_N$, and their corresponding acceleration $\rm a_{pred}$,
and their difference $\rm \Delta df$ are listed, together with the corresponding uncertainties.
}
\label{tab:DFpred}
\end{center}

\end{table}

\newpage
\ 

\end{document}